%% file: main.tex
\pdfoutput=1
\documentclass[11pt,a4paper]{article}

\usepackage[T1]{fontenc}
\usepackage[utf8]{inputenc}
\usepackage{lmodern}
\usepackage[margin=2.4cm]{geometry}
\usepackage{graphicx}
\usepackage{booktabs}
\usepackage{multirow}
\usepackage{amsmath}
\usepackage{microtype}
\usepackage[hidelinks]{hyperref}
\hypersetup{pdftitle={What reviewers call a Long Method is a length: a pre-registered test of coupling divergence as a god-unit signal, and what it settled instead},
  pdfauthor={Daniel Sagenschneider}, pdfsubject={Software engineering: empirical study of god-unit detectors}}
\usepackage{caption}
\input{generated/numbers}

\newcommand{\madz}{modified $z \geq 3.5$}
\newcommand{\fitwidth}[1]{\resizebox{\ifdim\width>\textwidth\textwidth\else\width\fi}{!}{#1}}
\newcommand{\sig}[1]{\texttt{#1}}
\newcommand{\Xs}{X$^*$}
\newcommand{\Dflow}{D$_{\text{flow}}$}
\newcommand{\dD}{$\Delta$D}
\newcommand{\dO}{$\Delta$O}

\title{\bfseries What reviewers call a Long Method is a length\\ A pre-registered test of coupling divergence as a god-unit signal, and what it settled instead}

\newcommand{\thedoi}{10.5281/zenodo.23749772}
\newcommand{\doiline}{\ifdefined\thedoi\quad\small\texttt{doi:\thedoi}\fi}

\author{Daniel Sagenschneider\\\small OfficeFloor\\\small\texttt{daniel@officefloor.net}\doiline}

\date{October 2026}

\begin{document}
\maketitle

\begin{abstract}
\noindent
A companion study found that a complexity outlier rule shows where
complexity concentrates but cannot tell a god unit from a complex
algorithm. This paper pre-registered a test of one way to tell them apart:
weight a unit's complexity by its \emph{coupling divergence}, the data it
hands on minus the data it receives, so that a leaf is set aside. Coupling
was measured by resolving every call in \mlcqMeasured{} MLCQ repositories,
\seedThreeMeasured{} fresh Java repositories and \fontanaSystems{} Qualitas
Corpus systems. The claim fails: the primary, the pooled MCC of the
weighted form minus complexity alone against MLCQ's human Blob, is
\primMicro{}, no difference resolved and underpowered, and the resolved
readings go against it. What the data settle is at method level. What
MLCQ's reviewers call a Long Method is an absolute length: a constant of
\lwLengthCutoff{} lines or more, chosen on the explore half, reaches MCC
\lwLengthMccConfirm{} on the confirm half, \lwLengthVsRule{} over the best
per-repository form; every reviewer with ten or more positives, on their
non-leaf reviews, has a best cutoff between \revCutoffMin{} and
\revCutoffMax{} lines; and nothing
measured inside a method, its calls, variables, data flow or the cost of
cutting it, refines the constant. The constant is a limit of human reading.
The next form of the construct is the cost of bringing a long method under
it, which is where the construct meets architecture.
\end{abstract}

\section{Introduction}

A companion study \cite{paperone} tested a god-class and god-method
detector with no raw-scale constant: a unit is flagged when its complexity
is a robust outlier for its own codebase, a modified z-score of 3.5 or more
on a scale fitted to that codebase \cite{iglewicz1993}. The rule found
where the complexity concentrates, and its extra flags were small classes
doing the work they were written to do. The difference between a god unit
and a complex algorithm, that study concluded, is in the coupling, and it
named this study as the test.

The construct tested here is \emph{coupling divergence}: what a unit hands
on minus what it receives. A unit that receives more than it sends is a
\emph{catchment}, a leaf that processes what it is given, where the
complexity of an algorithm belongs; one that sends more than it receives
is a \emph{junction}, and a unit that directs the data and also holds the
logic is what people call a god unit. The pre-registered claim was that
multiplying complexity by a factor that is zero for a catchment and grows
with the divergence of a junction recovers god classes and methods better
than complexity alone; its estimand, decision rule, sample and bootstrap
were committed before any coupling had been measured on the confirmatory
material.

The claim did not survive the test (Section~\ref{sec:claim}). What the
data settle instead is at method level: what MLCQ's reviewers call a Long
Method is an absolute length, \lwLengthCutoff{} lines or more, which a
per-repository rule is the wrong model for and which nothing measured
inside a method refines (Section~\ref{sec:length}). That constant is a
limit of human reading, and the construct's next form is the cost of
bringing a long method under it (Section~\ref{sec:disc}). The full
registered report is in the ancillary data.

\section{Background}

The relative outlier rule, the signal families and the oracles are the
companion study's \cite{paperone}. A unit is a method or a class (the
\texttt{.java} file); a method's lines are the code lines of its own body. Ten method signals in three
families (size, such as lines; structure, such as cyclomatic
\cite{mcCabe1976} and cognitive \cite{campbell2018} complexity; density,
such as cyclomatic density) give 31 class signals as a sum, a mean and a
maximum over the file's methods, plus the number of methods. The rule
standardises a signal within each repository by its median and median
absolute deviation and flags units at \madz{}.

Coupling as an ingredient of a god definition is old: Lanza and Marinescu's
God Class conjoins a weighted method count, low cohesion and access to
foreign data \cite{lanza2006, marinescu2004}, and Henry and Kafura's
information flow multiplies length by fan-in times fan-out
\cite{henry1981}, where size was later found to predict as well
\cite{kitchenham1990, elemam2001}, which is why a size-only weighting is
one of this paper's arms. Divergence differs from information flow in
subtracting what comes in from what goes out; its construct, a
\emph{sinew}, data that reaches the code that needs it only by riding the
call chain, descends from Yourdon and Constantine's tramp data
\cite{yourdon1979}. In Tempero and Ralph's terms \cite{tempero2026metrics} the
construct is the sinew and divergence one indicator of it; Arif, Kuutila
and Ralph found in-coupling and out-coupling loading as separate constructs
\cite{arif2026}, which is why the outgoing side is read on its own.
Absolute thresholds on length are older than the relative rule: Lanza and
Marinescu's threshold tables \cite{lanza2006} and Alves, Ypma and Visser's
benchmark-derived thresholds \cite{alves2010} fix a constant from a corpus
and apply it to every system, the model this paper's method-level result
returns to.

The human oracle is MLCQ \cite{mlcq2020}, professional developers' reviews
of Java open-source classes and methods, with the companion study's
strict-majority aggregation and its explore and confirm halves frozen
before any repository was measured; classifiers trained on it reach a
median MCC of about 0.51 for Blob and 0.81 for Long Method \cite{mlcq2023},
a reference point since the rules here see no label. Fontana and
colleagues' God Class instances on the Qualitas Corpus \cite{fontana2016}
are a second, balanced and extreme, human oracle.

\section{The registered test and its result}
\label{sec:claim}

\paragraph{The claim as registered.} For a method with $p$ parameters,
under the primary \emph{flow} definition, $\mathit{in} = 1 + p$ and
$\mathit{out} = \sum_c (1 + a(c))$ over its calls resolved to the
repository's own sources, each passing $a(c)$ arguments; a call to the JDK
or a dependency counts for nothing. The divergence is $d = \mathit{out} -
\mathit{in}$; a class's $\mathit{out}$ sums its methods' calls that leave
the file and its $\mathit{in}$ the parameters of its externally callable
methods. The divergence
factor is $f_D = \max(d + 1, 0)$, the product of an exemption $[d \geq 0]$
and an amplification $\max(d + 1, 1)$; the outgoing factor $f_O =
\mathit{out} + 1$ and the size null $f_N$, the class's number of methods,
run beside it, each scored by the same rule on every unit of the
repository, exempt units as zeros. The base signal \Xs{} was chosen on the
MLCQ explore half by complexity alone, as the signal whose flags maximise
the pooled MCC against human Blob. The estimand is the pooled MCC of
\Xs$\cdot$\Dflow{} minus \Xs{}, both by \madz{}, against MLCQ human Blob
over the reviewed classes of the confirm repositories, with a 95\%
percentile bootstrap over repositories (\bootB{} resamples, seed
\bootSeed{}): \emph{refines} on a lower bound above 0, \emph{worse} on an
upper bound below 0, otherwise \emph{no difference resolved}, and
\emph{underpowered} below a gate sized on the explore half. A consistency analysis then asked, for every oracle and signal family,
whether the divergence factor (\dD) and the outgoing factor (\dO) improve
on complexity alone, with four registered predictions and a verdict rule. The oracles are MLCQ's reviews on \mlcqMeasured{} measured
repositories (\mlcqBlobReviewed{} reviewed classes with \mlcqBlob{} Blobs,
\mlcqLongReviewed{} reviewed methods with \mlcqLong{} Long Methods, both
halves), five tool definitions on \seedThreeMeasured{} fresh Java
repositories, and Fontana's \fontanaInstances{} God Class instances on
\fontanaSystems{} Qualitas Corpus systems. Coupling is a JavaParser
measurer with its symbol solver, resolving against the repository's sources
and the JDK with no build; \callsMlcqUnresolved{} of confirm-half calls
resolve to nothing, almost always a dependency's type, so $\mathit{out}$
reads low.

\paragraph{The result.} Complexity alone chose \sig{\xClass} as \Xs{}
(explore MCC \xClassMcc{}); the gate asked for \gateRequired{} confirm
repositories with a reviewed Blob and the confirm half carries
\primGateReposWithBlob{}, so the primary is underpowered by the plan's rule
whatever it reads. On the confirm half complexity alone reaches MCC
\primXMcc{} and the weighted form \primDMcc{}; the paired difference is
\textbf{\primMicro{}}, verdict \textbf{\primVerdict}. The hypothesised
shape is inverted, recall up and precision down, and threshold-free the
weighted form ranks Blobs worse than complexity alone (AUC \primAucD{}
against \primAucX{}, \primAucDiff{}). Against the size null, a weighting by
the number of methods that carries no coupling, the coupling factor is
\textbf{\primDvsNMicro{}}: the rival explanation the plan recorded, that a
Blob is a big class with one complex method, beats the construct by a
resolved margin. The exemption alone has support in the labels: among the units \Xs{}
flags, a catchment is called a god unit less often by two of the three
human oracles (MLCQ Long Method \leafMethodDiff{}; Fontana
\leafFontanaReading{}; MLCQ Blob \leafReading{}), but the amplification
the factor adds costs more than the exemption gains. The construct's
verdict by the registered rule is \textbf{\consVerdict}: the outgoing
factor improves on complexity alone against human Blob in the peak and
density families (\dO{} \pOneOPeak{} and \pOneODensity{}) while subtracting
what comes in costs agreement in every family of every class-level oracle;
read threshold-free, post hoc, even that gain is a flag rate on MLCQ. At
method level the factor is \methMicro{} against Long Method, and no
secondary, sensitivity or tool oracle in the ancillary data changes this.

\paragraph{Deviations from the plan.} The plan was amended ten times after
registration, each amendment a commit stating what had been seen: eight
before any confirm or seed-3 repository had been measured, the ninth after
measurement but before any result had been read, and the tenth correcting
a fault found after the confirmatory results had been read. The harness
had attached a human oracle to a repository only when it held a positive
unit, so repositories whose reviewed units are all negatives never entered
the population; the fault was found after a confirmatory analysis had been
read by the author. The fix has no
free parameter; the explore-half choices and the analysis were made again
by the registered procedure, the earlier files set aside with their
digests. It restored \restoredBlobRepos{} confirm repositories and
\restoredBlobUnits{} reviewed classes to the Blob population; on the
set-aside analysis the primary read \priorPrimMicro{}, the same verdict.
The gate was not met and MLCQ has no more repositories, so a difference of
0.10 could have been missed. The full account is in the ancillary data.

\section{What the reviewers call a Long Method is a length}
\label{sec:length}

Everything in this section was computed after every oracle had been read,
by scripts in the ancillary data with their outputs and digests; it is
exploration and changes nothing above. The registered analysis had already
shown that length ranks Long Methods best of every form, better than the
registered cognitive signal, and that a gate on any internal call only
zeroes the short trivial methods and so recalibrates the per-repository
constant (percentile AUC against tokens
\matchedGatePctVsTokensConfirm{} on confirm).

\begin{table}[t]
\centering
\caption{Post hoc: MLCQ's \revLongReviews{} individual Long Method reviews
by \revReviewers{} reviewers on measured units (both halves), by the
method's length and whether it is a leaf (no internal call): the share
calling it a Long Method at any severity.}
\label{tab:reviewers}
\small
\setlength{\tabcolsep}{4pt}
\fitwidth{\input{generated/tab_reviewers}}
\end{table}

\paragraph{What the reviewers judge.} MLCQ keeps every review
(Table~\ref{tab:reviewers}). The share of non-leaf reviews calling a method
a Long Method is \revNonLeafTenRate{} up to 10 lines,
\revNonLeafTwentyFiveRate{} at 11 to 25, \revNonLeafFiftyRate{} at 26 to 50
and \revNonLeafHundredRate{} at 51 to 100: a step at about twenty-five
lines. Each of the \revCutoffReviewers{} reviewers with ten or more
positives has, on their non-leaf reviews, a best length cutoff between
\revCutoffMin{} and \revCutoffMax{} lines (median \revCutoffMedian{}), an
absolute constant
carried from reviewer to reviewer, which is why a cutoff tuned on MLCQ beat
the per-repository rule in the companion study and why no per-repository
form here can reach it. At equal length a leaf is called
a Long Method about as often as a non-leaf (\revLeafFiftyRate{} against
\revNonLeafFiftyRate{} at 26 to 50 lines; over all long methods
\revLongLeafMinusNonLeaf{}). The exemption the labels supported above is
narrower than the construct: among the methods cognitive complexity flags,
leaves are called Long Method less often (\revFlaggedLeafRate{} against
\revFlaggedRestRate{}) at the same median length, so it is the branchy
leaf reviewers decline to call a Long Method, not a leaf as such.

\paragraph{The constant, read once.} The constant the reviews imply was
read as the plan reads a cutoff: chosen on the explore half among
\lwNlocGridCount{} absolute cutoffs on lines, \lwNlocGrid{}, round numbers
with the companion study's tuned cutoff among them, and read on the confirm
half (\lwConfirmRepos{} repositories, \lwConfirmUnits{} reviewed methods,
\lwConfirmPositives{} positives). On the explore half the cutoffs read
\lwExploreTwentyMcc{} at 20, \lwExploreChosenMcc{} at \lwLengthCutoff{} and
\lwExploreThirtyMcc{} at 30, so the cutoff chosen is \lwLengthCutoff{} lines
or more. On confirm
it flags \lwLengthFlagged{} methods with \lwLengthHits{} hits, MCC
\lwLengthMccConfirm{} (precision \lwLengthPrecision{}, recall
\lwLengthRecall{}), \lwLengthVsRule{} over the best per-repository form,
the registered rule on length gated on any internal call. The MCC is flat
from 20 to 30 lines (\lwNlocTwentyMcc{} at 20, \lwNlocThirtyMcc{} at 30)
and has fallen by 40 (\lwNlocFortyMcc{}), so the constant is a mid-twenties
one rather than a sharp number.

\paragraph{What was tried against it.} Two further post hoc readings on
the same half asked whether anything inside a method refines that
constant. The first scored \lwHeldOutForms{} forms built from the method's
variables (the live width a reader holds at a line, the counts of locals,
parameters and fields, and the data flow from inputs to calls) by the same
procedure; every one is resolved below length (\lwFormsBelowLength{} of
\lwHeldOutForms{}), the best at \lwBestFormMcc{}, \lwBestFormVsLength{}.
The second scored the cost of cutting a long method into pieces under the
constant, the parameters and return values the pieces would exchange; its
cheapest-seam form reaches \splitCostMcc{}, \splitCostVsLength{} against
the constant, and among the \splitWithinN{} long methods, the only ground
it can gain, it ranks the \splitWithinPositives{} positives no better than
length does (AUC \splitCostAucWithin{} against \splitNlocAucWithin{},
\splitCostAucVsNloc{}).
Nothing in a long method's body, as these measures see it, says which of
the methods length flags the reviewers passed.

\section{Discussion}
\label{sec:disc}

\paragraph{Where the construct fails.} The factor was expected to raise
precision by setting complex leaves aside, and what a unit receives was
expected to matter as well as what it sends; neither is seen, and the
exemption the labels do support is narrowed by the reviews to the branchy
leaves complexity flags. A factor of $d + 1$ is unbounded and, for a class,
grows with the calls the class makes, so the weighted signal is dominated
by the classes that call most whatever their complexity; and a class with
a wide callable surface is one that other code uses, so subtracting its
surface pushes the hubs people call Blobs towards the catchment side, the
opposite of the construct's reading. Divergence as an indicator of the sinew fails
convergent validity \cite{tempero2026metrics} against three human
definitions of a god unit; and the human Blob is weakly predicted by
anything (complexity alone \primXMcc{}, trained classifiers 0.51
\cite{mlcq2023}), so part of the ceiling is the oracle.

\paragraph{What a Long Method is.} At method level the reviews settle more
than the forms do. What MLCQ's reviewers label a Long Method is a method of
\lwLengthCutoff{} lines or more, leaf or not, each reviewer read on enough
methods within a few lines of the others whatever the repository. The right
detector for that label is a constant on length, which the companion
study's tuned cutoff already reaches at 0.785 \cite{paperone} and which
reaches \lwLengthMccConfirm{} here; the per-repository rule is the wrong
model for it, and neither the registered construct nor anything read after
it inside the method refines it. What remains to register is whether the
branchy-leaf exemption the reviews show adds anything to the constant, and
whether the per-repository rule has a place on code whose scale has moved,
the purpose the plan stated and this data cannot test.

\paragraph{The next form of the sinew.} The constant is a limit of human
reading, which code that agents write and maintain need not respect, so
the construct's next form at method level is not a detector but a cost:
what it takes to bring a long method under the constant, counted as what
the pieces must exchange across each seam. That cost depends on who calls the pieces. When the remainder calls
them, every value that crosses is a parameter or the one return; when a
framework calls them in sequence and supplies what each declares, a value
handed on unchanged costs nothing. The difference between the two readings
of the same cut is what the architecture buys on the seam, and it is the
comparison to make between agent-written arms of one system under
different frameworks. Prior tools rank extract-method candidates within a
method \cite{tsantalis2011, haas2016}; none takes the length cap as a
constraint with a cost on the exchange under two calling conventions. That
instrument is built and is to be read descriptively before any claim is
registered for it.

\section{Threats to validity}
\label{sec:threats}

Divergence counts parameters and arguments, not what travels in them, and
a class's inputs count the parameters of its callable methods, so a wide
interface reads as receiving much, the likely failure of the sign. No
oracle names a complex algorithm; every oracle names a god unit, so the
title's question is answered from that side only. Unresolved calls make $\mathit{out}$ a lower bound, and injected
collaborators are invisible to static parsing. The study is Java only, and
\mlcqOwnerTopShare{} of MLCQ's confirm half sits in ten owners. The primary
is underpowered by its own gate, no secondary is corrected for
multiplicity, and the bootstraps are coarse at the tails. The post hoc readings of Section~\ref{sec:length} followed the constant
onto the confirm half, so that half is a hold-out for the choice among the
\lwNlocGridCount{} cutoffs alone, and the grid itself was drawn after the
reviews of both halves had been read. The plan's stated purpose, a definition
that moves with the codebase as agent-maintained code changes what a normal
level of complexity is, was not tested: no agent-written corpus with human
labels exists at the scale a pre-registered test needs.

\section{Conclusion}

Weighting complexity by coupling divergence does not recover the units
people call god units better than complexity alone; the primary resolves
no difference and is underpowered, and the resolved readings go the other
way. What the data settle is at method level: MLCQ's reviewers call a
method a Long Method at \lwLengthCutoff{} lines or more, leaf or not, a
constant on length (MCC \lwLengthMccConfirm{} on the confirm half) is the
detector that matches them, and nothing read inside a method refines it.
The constant is a limit of human reading; the form of sinew left to build
is the cost of the cut.

\paragraph{Data availability.} Every number here is produced by a script
from the harness's analysis files, and the table is a generated fragment.
The ancillary data holds the full registered report, the confirmatory
analysis and the two set aside with their digests, the account of every
deviation, the explore-half decisions, run records, pinned samples,
converted labels, and the post hoc scripts with their outputs and digests;
its README names every repository, commit and command.

\section*{Competing interests and funding}

The author created and maintains OfficeFloor, a framework whose design
rests on taking data off the control flow, from which the sinew construct
comes; a positive result would have supported the framework's premise, and
the result is negative. The population fault (Section~\ref{sec:claim}) was found after the
confirmatory results had been read; the fix is the one the plan's text
dictates, and every headline reading before and after it is reported. The
plan, seed, decision rule and gate were fixed before the data existed; a
replication needs no cooperation from the author, whose companion study,
sampler and measurer these are. The work received no external funding.

\small
\let\oldthebibliography\thebibliography
\renewcommand{\thebibliography}[1]{\oldthebibliography{#1}\setlength{\itemsep}{2pt}}

\end{document}

%% file: generated/numbers.tex
\newcommand{\bootB}{2{,}000}
\newcommand{\bootSeed}{12345}

\newcommand{\mlcqBlobReviewed}{2{,}334}
\newcommand{\mlcqBlob}{254}
\newcommand{\mlcqLongReviewed}{2{,}430}
\newcommand{\mlcqLong}{276}

\newcommand{\mlcqMeasured}{520}

\newcommand{\seedThreeMeasured}{500}

\newcommand{\fontanaSystems}{73}
\newcommand{\fontanaInstances}{420}

\newcommand{\callsMlcqUnresolved}{24.0\%}

\newcommand{\mlcqOwnerTopShare}{50.8\%}

\newcommand{\xClass}{nloc\_max}
\newcommand{\xClassMcc}{0.287}

\newcommand{\gateRequired}{72}

\newcommand{\primMicro}{$-$0.041 [$-$0.130, +0.054]}

\newcommand{\primVerdict}{no difference resolved (underpowered)}

\newcommand{\primGateReposWithBlob}{66}
\newcommand{\primXMcc}{0.187}

\newcommand{\primDMcc}{0.146}

\newcommand{\primAucX}{0.768}
\newcommand{\primAucD}{0.631}
\newcommand{\primAucDiff}{$-$0.137 [$-$0.195, $-$0.080]}

\newcommand{\primDvsNMicro}{$-$0.091 [$-$0.175, $-$0.006]}

\newcommand{\leafReading}{unresolved}

\newcommand{\leafMethodDiff}{$-$0.304 [$-$0.541, $-$0.082]}

\newcommand{\leafFontanaReading}{supported}

\newcommand{\consVerdict}{outgoing data only}

\newcommand{\pOneOPeak}{+0.125 [+0.054, +0.195]}
\newcommand{\pOneODensity}{+0.151 [+0.076, +0.227]}

\newcommand{\methMicro}{$-$0.092 [$-$0.188, +0.010]}

\newcommand{\priorPrimMicro}{$-$0.075 [$-$0.191, +0.045]}

\newcommand{\restoredBlobRepos}{141}
\newcommand{\restoredBlobUnits}{386}

\newcommand{\revLongReviews}{3{,}054}

\newcommand{\revReviewers}{26}

\newcommand{\revNonLeafTenRate}{1.0\%}

\newcommand{\revNonLeafTwentyFiveRate}{24.3\%}

\newcommand{\revLeafFiftyRate}{80.0\%}
\newcommand{\revNonLeafFiftyRate}{81.6\%}

\newcommand{\revNonLeafHundredRate}{94.0\%}

\newcommand{\revLongLeafMinusNonLeaf}{$-$0.033 [$-$0.141, +0.045]}

\newcommand{\revCutoffReviewers}{10}
\newcommand{\revCutoffMin}{17}
\newcommand{\revCutoffMax}{34}
\newcommand{\revCutoffMedian}{21}

\newcommand{\revFlaggedLeafRate}{63.0\%}
\newcommand{\revFlaggedRestRate}{83.3\%}

\newcommand{\matchedGatePctVsTokensConfirm}{+0.005 [$-$0.002, +0.013]}

\newcommand{\lwLengthCutoff}{26}
\newcommand{\lwLengthMccConfirm}{0.777}
\newcommand{\lwLengthFlagged}{131}
\newcommand{\lwLengthHits}{103}
\newcommand{\lwLengthPrecision}{0.79}
\newcommand{\lwLengthRecall}{0.82}
\newcommand{\lwLengthVsRule}{+0.080 [+0.016, +0.145]}
\newcommand{\lwHeldOutForms}{34}
\newcommand{\lwBestFormMcc}{0.673}
\newcommand{\lwBestFormVsLength}{$-$0.104 [$-$0.174, $-$0.035]}

\newcommand{\lwFormsBelowLength}{34}

\newcommand{\lwNlocTwentyMcc}{0.761}
\newcommand{\lwNlocThirtyMcc}{0.778}
\newcommand{\lwNlocFortyMcc}{0.644}
\newcommand{\lwConfirmRepos}{202}
\newcommand{\lwConfirmUnits}{1{,}033}
\newcommand{\lwConfirmPositives}{125}
\newcommand{\lwNlocGrid}{10, 15, 20, 26, 30, 40, 50, 75 and 100}
\newcommand{\lwNlocGridCount}{9}
\newcommand{\lwExploreTwentyMcc}{0.710}
\newcommand{\lwExploreThirtyMcc}{0.681}
\newcommand{\lwExploreChosenMcc}{0.715}

\newcommand{\splitCostMcc}{0.764}
\newcommand{\splitCostVsLength}{$-$0.013 [$-$0.043, +0.015]}

\newcommand{\splitWithinN}{130}
\newcommand{\splitWithinPositives}{103}
\newcommand{\splitNlocAucWithin}{0.789 [0.693, 0.879]}
\newcommand{\splitCostAucWithin}{0.707 [0.591, 0.828]}
\newcommand{\splitCostAucVsNloc}{$-$0.082 [$-$0.198, +0.036]}

%% file: generated/tab_reviewers.tex
\begin{tabular}{lrrrr}
\toprule
Lines of the method & Leaf reviews & called Long Method & Non-leaf reviews & called Long Method \\
\midrule
1-10 & 770 & 0.4\% & 823 & 1.0\% \\
11-25 & 230 & 17.4\% & 571 & 24.3\% \\
26-50 & 80 & 80.0\% & 365 & 81.6\% \\
51-100 & 17 & 94.1\% & 151 & 94.0\% \\
101+ & 7 & 85.7\% & 40 & 95.0\% \\
\bottomrule
\end{tabular}